\documentclass[runningheads]{llncs}
\usepackage{graphicx}
\usepackage{amsmath,amssymb} 
\usepackage{color}
\usepackage{multirow}
\usepackage{tabularx,ragged2e}
\usepackage{booktabs}
\usepackage{subcaption}
\begin{document}
\pagestyle{headings}
\mainmatter

\title{Deep Slice Interpolation via Marginal Super-Resolution, Fusion and Refinement} 



\author{Cheng Peng$^{1}$, Wei-An Lin$^{1}$, Haofu Liao$^{2}$, Rama Chellappa$^{1}$, S. Kevin Zhou$^{3,4}$}
\institute{$^{1}$The University of Maryland, College Park $^{2}$University of Rochester $^{3}$Chinese Academy of Sciences $^{4}$Peng Cheng Laboratory, Shenzhen}

\maketitle

\begin{abstract}
We propose a marginal super-resolution (MSR) approach based on 2D convolutional neural networks (CNNs) for interpolating an anisotropic brain magnetic resonance scan  along the highly under-sampled direction, which is assumed to axial without loss of generality. Previous methods for slice interpolation only consider data from pairs of adjacent 2D slices. The possibility of fusing information from the direction orthogonal to the 2D slices remains unexplored. Our approach performs MSR in both sagittal and coronal directions, which provides an initial estimate for slice interpolation. The interpolated slices are then fused and refined in the axial direction for improved consistency. Since MSR consists of only 2D operations, it is more feasible in terms of GPU memory consumption and requires fewer training samples compared to 3D CNNs. Our experiments demonstrate that the proposed method outperforms traditional linear interpolation and baseline 2D/3D CNN-based approaches. We conclude by showcasing the method's practical utility in estimating brain volumes from under-sampled brain MR scans through semantic segmentation. 
\end{abstract}

\section{Introduction}

Magnetic resonance imaging (MRI) has been one of prevailing gold standards for diagnostic purposes. It is not only non-invasive, but also better at targeting different human tissues with specific contrasts that reveal the underlying anatomy. The main disadvantage of MRI compared to other medical imaging modalities (e.g. computed tomography, or CT) is its long acquisition time, which is governed by the duration of the frequency signals to be emitted by atoms and sampled by the machine. There has been a long history of studies on accelerating the MRI sampling process \cite{DBLP:journals/tmi/RavishankarB11,lustig2007sparse,DBLP:conf/cvpr/MaYZC08/CSMRI,DBLP:journals/tmi/SchlemperCHPR18} by undersampling in the 2D k-space during acquisition; however, only a relatively small number of studies \cite{DBLP:journals/tmi/GoshtasbyTA92,DBLP:journals/tmi/GreveraU96,DBLP:journals/tmi/LeeW00,DBLP:journals/tmi/PenneySRVN04} focused on interpolating between the sampled slices.

\begin{figure}[htb]
\centering
\includegraphics[width=0.7\textwidth]{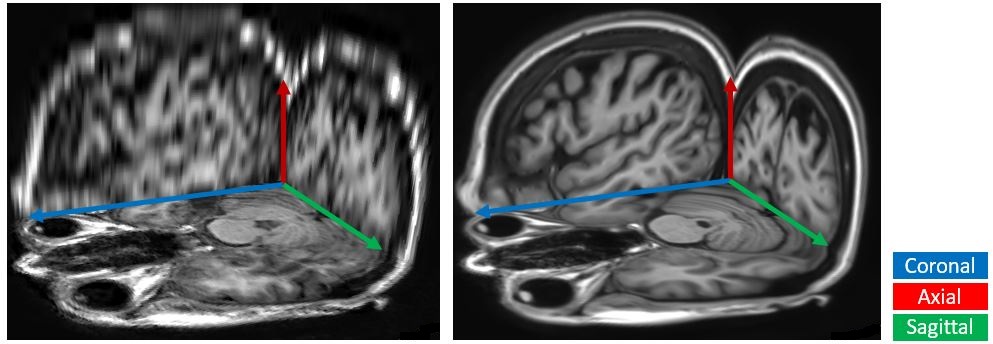}
\caption{The axial, coronal, and sagittal views of an anisotropic MR volume are fitted to isotropic resolution through (Left) linear interpolation and (Right) our proposed slice-interpolation method.}
\label{fig:intro}
\vspace{-1em}
\end{figure}

In practice, 
most MR volumes are taken anisotropically with a high resolution within slices and a sparse resolution between slices. For example, Fig. \ref{fig:intro} shows a brain MR scan whose axial direction is sparsely sampled. As a result, image quality suffers when viewing from coronal and sagittal directions. 

It is desirable to have a consistent resolution across all dimensions, both for visualization and for medical analysis tasks such as brain volume estimation. Traditionally, slice interpolation has been done with two groups of methods: intensity-based and deformation-based methods. Linear and cubic spline interpolation methods are classic examples of intensity-based methods that directly perform the interpolation based on the intensity of the adjacent slices. Deformation-based methods estimate deformation fields between adjacent slices, and then interpolate in-between pixels based on the estimated fields. However, these methods require that adjacent MR slices contain similar anatomical structures. That is, the structural change must be sufficiently small so that a dense pixel correspondence can be established between adjacent slices. When the anatomical variation between slices is significant, more sophisticated modeling approach is needed.

Recently, deep convolutional neural networks (DCNNs) have been outperforming traditional approaches on medical image analysis due to their ability to model complex variations within data \cite{DBLP:journals/tmi/SchlemperCHPR18,DBLP:conf/miccai/RonnebergerFB15,DBLP:journals/corr/abs-1711-08580}. For slice interpolation, DCNNs can be applied to learn a mapping from an anisotropic MR to isotropic. However, directly addressing the task in 3D is challenging due to the high memory consumption of 3D networks. In this work, we break down the task of 3D slice interpolation into a sequence of 2D problems to produce anatomically consistent slice interpolations while being memory-feasible. Specifically, we propose a novel \emph{marginal super-resolution} to super-resolve isotropic views in the sagittal and coronal directions by a 2D CNN. The interpolation along the axial direction can be estimated by a \emph{fusion} of the isotropic saggital and coronal views. Finally, the interpolated slices are processed to recover more details via \emph{refinement}.


Our main contributions can be summarized as follows:
\begin{itemize}
    \item We propose a novel marginal super-resolution approach to break down the 3D slice interpolation problem into several 2D problems, which is more feasible in terms of GPU memory consumption and the amount of data available for training.
    \item We propose a two-view fusion approach to incorporate the 3D anatomical structure. The interpolated slices after fusion achieve high structural consistency. The final refinement further recovers fine details.
    \item We perform extensive evaluations on a large-scale MR dataset, and show that the proposed method outperforms all the competing CNN models, including 3D CNNs, in terms of quantitative measurement, visual quality, and brain matter segmentation. 
\end{itemize}

\section{Related Work}


\subsubsection{Traditional slice interpolation methods.}
Early work on interpolating volumetric medical data dates back to 1992, when Goshtasby et al.~\cite{DBLP:journals/tmi/GoshtasbyTA92} proposed to leverage the small and gradual anatomic differences between consecutive slices, and find correspondence between pixels by searching through small neighborhoods. A slew of methods were proposed in the subsequent years, focusing on finding more accurate deformation fields, including shape-based methods~\cite{DBLP:journals/tmi/GreveraU96}, morphology-based methods~\cite{DBLP:journals/tmi/LeeW00}, registration-based methods~\cite{DBLP:journals/tmi/PenneySRVN04}, etc. Linear interpolation can be regarded as a special example, which essentially assumes no deformation between slices.

An important assumption made in the above-mentioned methods is that adjacent slices contain similar anatomical structures, i.e., the changes in the structures have to be sufficiently small such that a dense correspondence can be found between two slices. This assumption largely limits the applicability of slice interpolation methods especially when slices are sparsely sampled. Furthermore, these methods did not utilize the information outside the two adjacent slices.

\subsubsection{Learning based super-resolution methods.}
Slice interpolation can be viewed as a special case of 3D super-resolution. Here we review the literatures of 2D Single Image Super-Resolution (SISR), especially those approaches based on CNNs. 
Dong et al. \cite{DBLP:journals/corr/DongLHT15} first proposed SRCNN, which learns a mapping that optimally transforms low-resolution (LR) images to high-resolution (HR) images. Many subsequent studies explored strategies to improve SISR such as using deeper archtectures and weight-sharing \cite{DBLP:journals/corr/KimLL15b,DBLP:conf/cvpr/ZhangZGZ17,DBLP:conf/cvpr/KimLL16}. However, these methods require bilinear upsampling as a pre-processing step, which drastically increases computational complexity \cite{DBLP:journals/corr/DongLT16}. To address this issue, Dong et al. \cite{DBLP:journals/corr/DongLT16} proposed to apply deconvolution layers for LR image to be directly upsampled to finer resolution. Furthermore, many studies have shown that residual learning provided better performance in SISR \cite{DBLP:journals/corr/LimSKNL17,DBLP:conf/cvpr/LedigTHCCAATTWS17,DBLP:journals/corr/abs-1802-08797}. Specifically, Zhang et al. \cite{DBLP:journals/corr/abs-1802-08797} incorporated both residual learning and dense blocks \cite{DBLP:journals/corr/HuangLW16a}, and introduced Residual Dense Blocks (RDB) to allow for all layers of features to be seen directly by other layers, achieving state-of-the-art performance. 

Generative Adversarial Networks (GAN) \cite{DBLP:journals/corr/GoodfellowPMXWOCB14} have also been incorporated in SISR to improve the visual quality of the generated images. Ledig et al. pointed out that training SISR networks solely by $L_1$ or $L_2$ loss intrinsically leads to blurry estimations, and proposed SRGAN \cite{DBLP:conf/cvpr/LedigTHCCAATTWS17}, which generated much sharper and realistic images compared to other approaches, despite having lower peak signal to noise ratios.

Though available computation capacity has been increasing, 3D CNNs are still limited by memory capacity due to a considerable increase in the size of network parameters and input data. A common compromise is to extract small patches from 3D volume to reduce the input size \cite{DBLP:journals/corr/abs-1803-01417}; however, this also limits the effective receptive field of the network. In practice, 3D CNNs are also limited by the amount of training data to ensure generalization.

\section{Problem formulation and baseline CNN approaches}
Let $I(x,y,z) \in \mathbb{R}^{N\times N\times N}$ denote an isotropic MR volume. By convention, we refer the $x$ axis as the ``sagittal'' axis, the $y$ axis as the ``coronal'' axis, and the $z$ axis as the ``axial'' axis. Accordingly, there are three types of slices:
\begin{itemize}
\item the sagittal slice for a given $x$: $I^{x}(y,z)= I(x,y,z), \forall x$;
\item the coronal slice for a given $y$: $I^{y}(x,z)= I(x,y,z), \forall y$;
\item the axial slice for a given $z$: $I^{z}(x,y) = I(x,y,z), \forall z$.
\end{itemize}
We also define a slab of $s$ slices, say along the $x$ axis, as
\begin{align}
    \mathbb{I}^{x,s}&=\left\{I^{x+l}(y,z) \bigg | l=-\frac{s-1}{2}, \ldots , 0 , \ldots ,\frac{s-1}{2}\right\}.    
\end{align}
$\mathbb{I}^{y,s}$ and $\mathbb{I}^{z,s}$ are defined similarly.
Without loss of generality, in this work we consider slice interpolation along the axial axis. From $I(x,y,z)$, the corresponding anisotropic MR volume is defined as 
\begin{align}
I_{\downarrow k}(x,y,z) = I(x,y,k \cdot z),     
\end{align}
where $k$ is the sparsity factor.
The {\it goal of slice interpolation} is to find a transformation  $\mathcal{T}\colon{\mathbb{R}^{N\times N\times \frac{N}{k}}}\to{\mathbb{R}^{N\times N\times N}}$ that can optimally transform $I_{\downarrow k}(x,y,z)$ back to $I(x,y,z)$. 

There are two possible baseline realizations of $\mathcal{T}$ using CNNs: 
\begin{itemize}
    \item {\bf 2D CNN.}
More in line with the traditional methods, a 2D CNN takes two adjacent slices $I^{z}_{\downarrow k}(x,y)$ and $I^{z+1}_{\downarrow k}(x,y)$ as inputs, and directly estimates the in-between missing slices. One major drawback of this approach is that a simple 2D CNN has limited capabilities of modeling the variations in highly anisotropic volumes.

\item {\bf 3D CNN.}
A 3D CNN is learned as a mapping from the sparsely sampled volume $I_{\downarrow k}(x,y,z)$ to a fully sampled volume $I(x,y,z)$. This straightforward approach, however, suffers from training memory issue and insufficient training data.
\end{itemize}
Below, we present our proposed algorithm that retains the advantages of the baseline CNN models discussed above while mitigating their disadvantages.

\section{The Proposed Algorithm}

We propose to break down the 3D slice interpolation problem into a series of 2D tasks, and interpolate the contextual information from all three anatomical views to achieve structurally consistent reconstruction and improved memory efficiency. The two stages are as follows:
\begin{itemize}
    \item Marginal super-resolution (MSR), where we provide high-quality estimates of the interpolated slices by extrapolating context from sagittal and coronal axes.
    \item Two-view Fusion and Refinement (TFR), where we fuse the estimations and further refine with information from the axial axis.
\end{itemize}

\begin{figure}[]
\centering
\includegraphics[width=0.8\textwidth,height=0.3\textwidth]{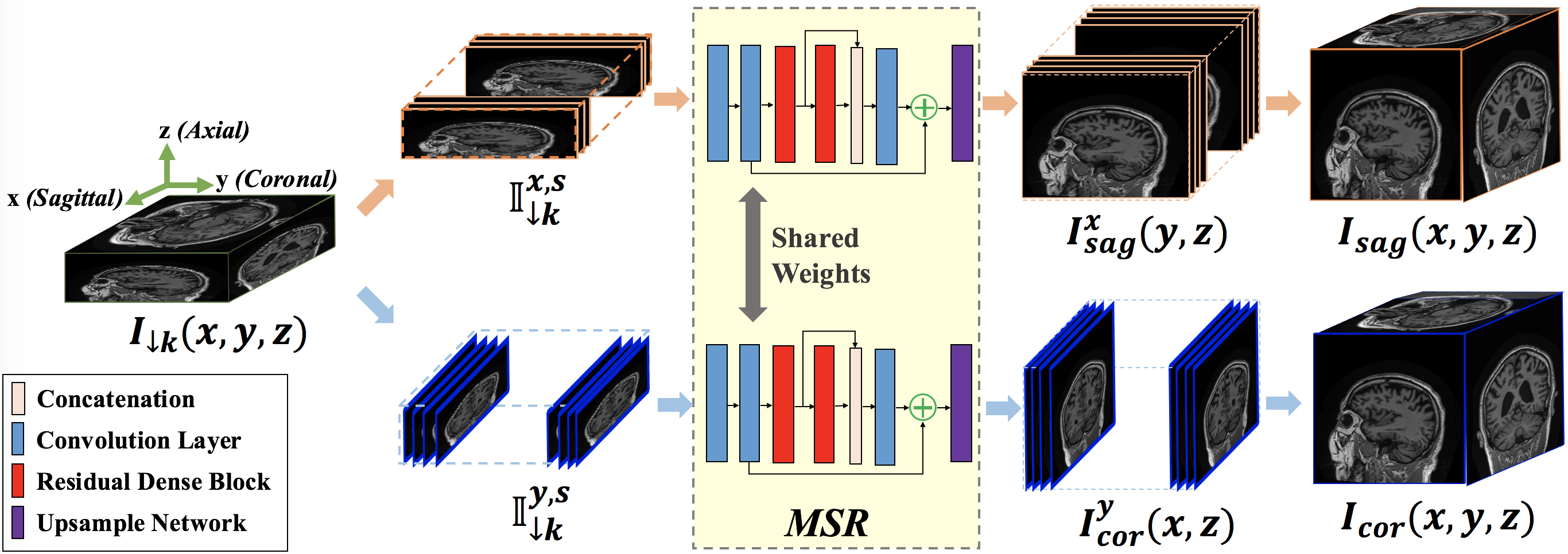}
\caption{Marginal Super-Resolution Pipeline.}\label{fig:MSR}
\vspace{-1em}
\end{figure}

\subsection{Marginal Super-Resolution} 


Fig. \ref{fig:MSR} demonstrates the pipeline of MSR. Given $I_{\downarrow k}(x,y,z)$, we view it as a sequence of 2D sagittal slices $I^{x}_{\downarrow k}(y,z)$ marginally from the sagittal axis. The same volume can also be treated as $I^{y}_{\downarrow k}(x,z)$ from the coronal axes. We make an observation that super-resolving $I^{x}_{\downarrow k}(y,z)$ to $I^{x}(y,z)$ and $I^{y}_{\downarrow k}(x,z)$ to $I^{y}(x,z)$ are equivalent to applying a sequence of 2D super-resolution along the $x$ axis and $y$ axis, respectively. Therefore, we apply a residual dense network (RDN) \cite{DBLP:journals/corr/abs-1802-08797} $\mathcal{M}_{\theta}$ to upsample $I^{x}_{\downarrow k}(y,z)$ and $I^{y}_{\downarrow k}(x,z)$ as follows: 
\begin{align}
    I^{x}_{sag}(y,z)=\mathcal{M}_{\theta}(\mathbb{I}^{x,s}_{\downarrow k}(y,z)),~I^{y}_{cor}(x,z)=\mathcal{M}_{\theta}(\mathbb{I}^{y,s}_{\downarrow k}(x,z)).
\end{align}
Notice that instead of super-resolving 2D slices independently, we propose to take a slab of $s$ slices as input and estimate a single SR output. Using a larger $s$ allows more context to be modelled. The MSR process is repeated for all $x$ and $y$.
Finally, the super-resolved slices can be reformatted as sagittally and coronally super-resolved volumes, $I_{sag}(x,y,z)$ and $I_{cor}(x,y,z)$, respectively. We apply the following $L_1$ loss to train the RDN:
\begin{align}
    \mathcal{L}_{MSR} &= \lVert \mathcal{M}_{\theta}(\mathbb{I}^{x,s}_{\downarrow k}) - I^{x}_{gt} \rVert_1 + \lVert \mathcal{M}_{\theta}(\mathbb{I}^{y,s}_{\downarrow k}) - I^{y}_{gt} \rVert_1,
\end{align}
where $I^{x}_{gt} = I^x(y, z)$ and $I^{y}_{gt} = I^y(x, z)$ in the isotropic MR volume.

From the axial perspective, $I_{sag}(x,y,z)$ and $I_{cor}(x,y,z)$ provide line-by-line estimations for the missing axial slices. However, since no constraint is enforced on the estimated axial slices, inconsistent interpolations lead to noticeable artifacts (See Section \ref{sec:ablation_study}).
We resolve this problem in the second TFR stage of the proposed pipeline. 

\subsection{Two-View Fusion and Refinement}
\begin{figure}[]
\centering
\includegraphics[width=0.8\textwidth]{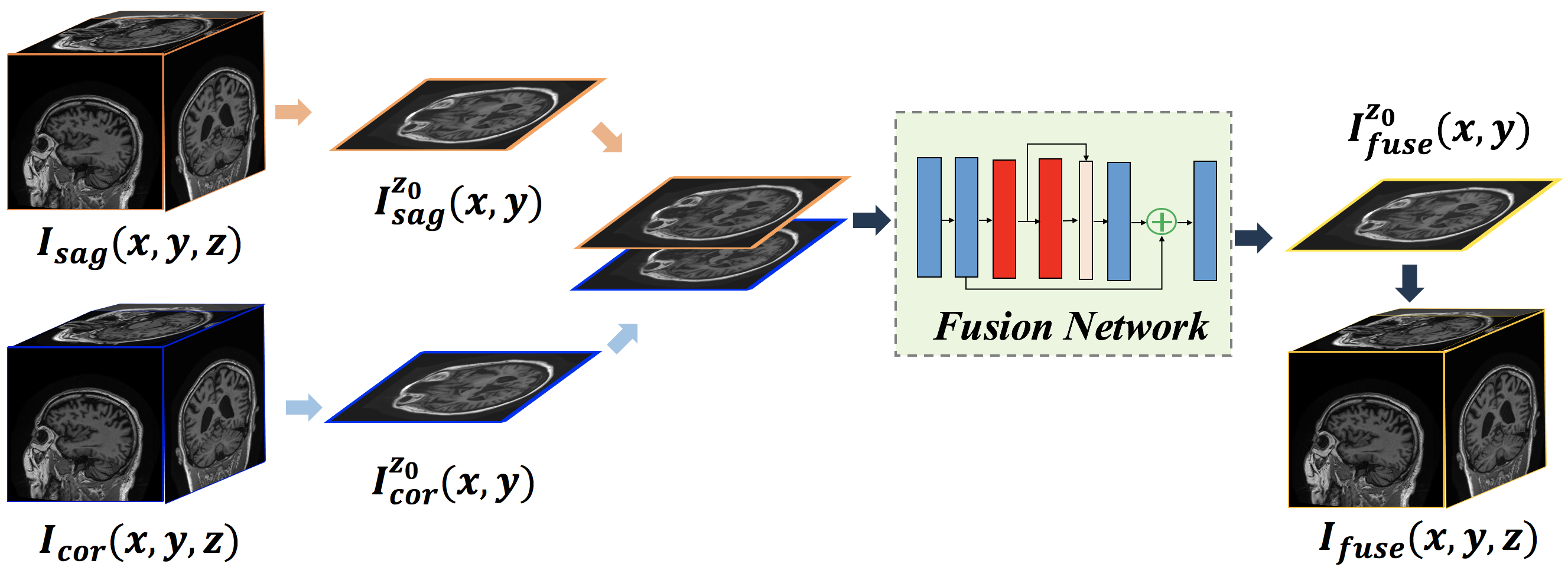}
\caption{Two-view Fusion Pipeline.}\label{fig:Fusion}
\vspace{-1.5em}
\end{figure}

The TFR stage is the counterpart of MSR which further improves the quality of slice interpolation by learning the structural variations along the axial direction. 

As shown in Fig. \ref{fig:Fusion}, we first resample the sagitally and coronally super-resovled volumes $I_{sag}(x,y,z)$ and $I_{cor}(x,y,z)$ from the axial direction to obtain $I_{sag}^{z}(x,y)$ and $I_{cor}^{z}(x,y)$, respectively. A fusion network $\mathcal{F_{\phi}}$ takes the two slices as inputs and combines information from the two views. The objective function for training the fusion network is:
\begin{align}
    \mathcal{L}_{fuse} &= \lVert I^{z}_{fuse}(x,y) - I^{z}_{gt} \rVert_1,
\end{align}
where $I^{z}_{fuse}(x,y)= \mathcal{F}_{\phi}(I_{sag}^{z}, I_{cor}^{z})$ is the output of the fusion network, and $I^{z}_{gt} = I^z(x, y)$ in the isotropic MR volume. After training, the fusion network is applied to all the \emph{interpolated} slices $\{I_{sag}^z ~|~ (z\mod k) \neq 0\}$ and $\{I_{cor}^z ~|~ (z \mod k) \neq 0\}$, yielding an MR volume $I_{fuse}(x,y,z)$.



\begin{figure}[htb]
\vspace{-0.5cm}
\centering
\includegraphics[width=0.8\textwidth]{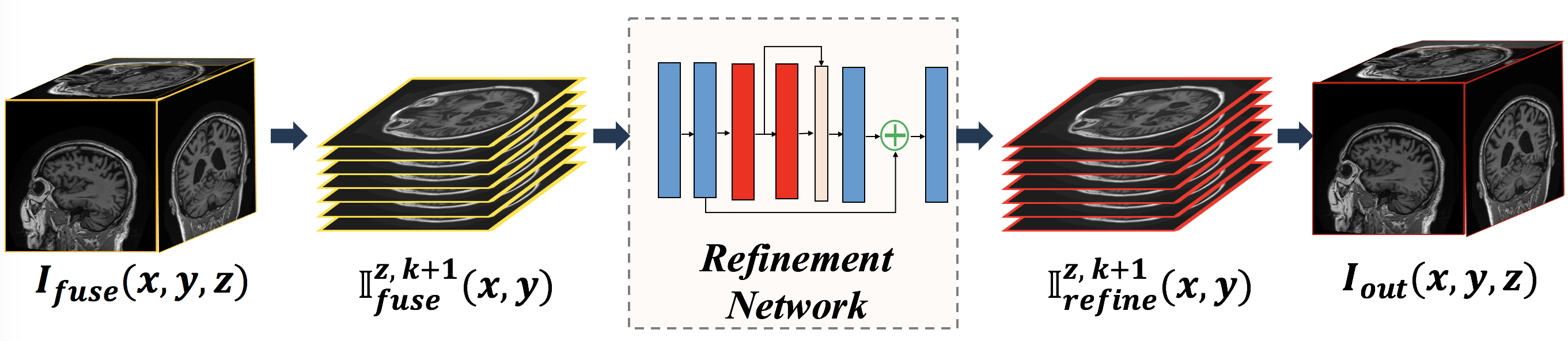}
\caption{Refinement Pipeline.}\label{fig:Refinement}
\vspace{-1em}
\end{figure}
After fusion, the interpolated slices already have visually pleasing qualities. Finally, to improve between-slice consistency along the axial axis, a refinement network $\mathcal{R}_{\psi}$ takes a slab of $k+1$ slices $\mathbb{I}_{fuse}^{z, k+1}$ as input and generates a consistent output slab $\mathbb{I}_{refine}^{z, k+1}$. The size is selected as $k+1$ to make sure the refinement network has information from one or two \emph{observed} slices. The pipeline is illustrated in Fig. \ref{fig:Refinement}. The loss function is given by:
\begin{align}
    \mathcal{L}_{refine} &= \lVert \mathbb{I}^{z, k+1}_{refine} - \mathbb{I}^{z, k+1}_{gt} \rVert_1.
\end{align}

\subsection{Comparison with baseline CNN approaches}
A 2D CNN estimates missing slices solely based on adjacent MR scans. In contrast, the proposed MSR and TFR take into account the full context from sagittal, coronal, and axial views, thus providing strong estimates of the in-between slices. A 3D CNN directly maps a sparsely sampled MR volume to a fully sampled MR volume. Due to memory limitation, a volume often needs to be divided into small patches during training, which limits the effective receptive field of 3D CNNs. In the proposed method, interpolation in 3D space is treated as a sequence of 2D operations, which ensures that the networks can be trained without relying on patches, thus allowing full contextual information to be captured. Furthermore, there are sufficient samples to train 2D CNNs, which mitigates the problem of overfitting issue that plagues 3D CNNs.


\section{Experiments}

\subsection{Settings}
\subsubsection{Implementation Details}
 We implement the proposed framework using PyTorch\footnote{https://pytorch.org}. The RDN \cite{DBLP:journals/corr/abs-1802-08797} architecture with two RDBs are used as the building unit for our networks. For Fusion, Refinement, and baseline 2D CNN models, where the inputs and outputs have the same image size, we replace the upsampling network in RDN with one convolutional layer. The input to the MSR network has $s=3$. Note that due to memory constraint, 3D CNN only uses one RDB. We train the models with Adam
optimization, with a momentum of 0.5 and a learning rate of 0.0001, until they reach convergence.

\subsubsection{Dataset}
We employ 120 T1 MR brain scans from the publicly available Alzheimer's Disease Neuroimaging Initiative (ADNI) dataset. The MR scans are isotropically sampled at 1 mm $\times$ 1 mm $\times$ 1 mm, and zero-padded to $256\times256\times256$ pixels, ending up with 30,720 slices in each of sagittal, coronal, and axial directions.
We further down-sample the isotropic volumes by factors of $k=4$ and $k=8$, yielding $I_{\downarrow k}(x,y,z)$ of sizes $256\times256\times64$ and $256\times256\times32$, respectively. The data is split into training/validation/testing sets with 95/5/20 samples. Note that during test time, we only select slices that contain mostly brain tissues, the number of samples for each sparsity are presented in Table \ref{tab:Comparison}.

\subsubsection{Evaluation metrics}
We compare different slice interpolation approaches using two types of quantitative metrics. First, we use Peak Signal-to-Noise Ratio (PSNR) and Structured Similarity Index (SSIM) to measure low-level image quality. Second, we evaluate the quality of the interpolated slices through gray/white-matter segmentation. The segmentation network has a U-Net architecture, which is one of the winning models in MRBrainS challenge \cite{DBLP:journals/jdi/AkkusGHRE17}, and is trained on the OASIS dataset \cite{DBLP:journals/jocn/MarcusWPCMB07}. Dice Coefficient (DICE) and Hausdorff Distance (HD)\footnote{To reduce the effect of outliers, HD is calculated on the 90th percentile displacement.} between the segmentation maps of ground truth slices and generated slices are calculated. Due to the memory limitation of 3D CNN, we can at most super-resolve a limited region of $144\times 144 \times 256$ pixels during evaluation. For fair comparisons, the evaluation metrics are calculated over the same region across all methods.

\subsection{Quantitative Evaluations}
In this section, we evaluate the performance of our method and the baseline approaches. Quantitative comparisons are presented in Table \ref{tab:Comparison}. We can observe that all the three CNN based methods have higher PSNR and SSIM than the widely used linear interpolation. 3D CNN slightly outperforms 2D CNN in 4x sparsity, but performs worse in 8x sparsity. Among the three CNN methods, our method consistently outperforms 2D CNN and 3D CNN baselines. 

The performance gain in accurately segmenting gray and white matters is large from linear interpolation to baseline CNN-based methods. However, at 8x sparsity, the HD scores of linear interpolation are comparable with 2D CNN and 3D CNN, while our method outperforms these approaches by at least 10$\%$. This demonstrates the robustness of our method even at very high sparsity.

\newcolumntype{Y}{>{\centering\arraybackslash}X}

\begin{table*}[t!]
\small
\begin{tabularx}{\textwidth}{ |c|c| *{4}{Y|} }
\hline

\hline
Sparsity & Method & PSNR(dB) & SSIM & DICE & HD(90th pct.)  \\
&&&&GM/WM&GM/WM\\
\hline
\multirow{4}{0.5em}{4} &LI &26.39 & 0.8317 & 0.7716/0.7296 & 3.607/7.965  \\
 &2D CNN  &  31.24 & \underline{0.9313} & \underline{0.8813}/\underline{0.8334}  & 3.176/12.36 \\
 &3D CNN  &  \underline{31.34} & 0.9292 & 0.8536/0.8265 & \underline{2.898}/\underline{7.373} \\
 &Ours  &  {\bf 32.22} & {\bf 0.9441} & {\bf 0.9021}/{\bf 0.8593} & {\bf 2.494}/{\bf 6.240} \\ 
\hline
\multirow{4}{0.5em}{8} &LI& 23.45 & 0.7165 & 0.6611/0.6105 & 4.487/10.59\\
 &2D CNN  & \underline{27.88} & \underline{0.8444} & \underline{0.7783}/0.7425 & \underline{4.322}/12.84 \\
 &3D CNN  & 27.38 & 0.8390 & 0.7684/\underline{0.7468} & 4.583/\underline{9.017} \\
 &Ours  & {\bf28.87} & {\bf0.8808} & {\bf0.8189}/{\bf0.7828} & {\bf3.960}/{\bf8.127} \\  
\hline

\end{tabularx}
\vspace{2pt}
\caption{Quantitative evaluations for different slice interpolation approaches. For DICE and HD performance metrics, we present results on gray matter (GM)/white matter (WM) segmentation. The best results are in {\bf bold} and the second best \underline{underlined}.}
\label{tab:Comparison}
\vspace{-0.5cm}
\end{table*}

\newcolumntype{Y}{>{\centering\arraybackslash}X}

\begin{figure*}[t!]
\captionsetup[subfigure]{labelformat=empty}
\begin{tabularx}{\textwidth}{ |c| *{7}{Y}| }
\hline

Sparsity& $I^{z}_{\downarrow k}$ & LI & 2D CNN & 3D CNN & Ours & GT & $I^{z+1}_{\downarrow k}$ \\
\hline
\multirow{2}{0.5em}{4} &
\begin{subfigure}[t]{0.125\textwidth}
    \centering
    \includegraphics[width=\linewidth]{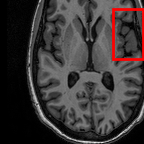}
\end{subfigure} &
\begin{subfigure}[t]{0.125\textwidth}
    \centering
    \includegraphics[width=\linewidth]{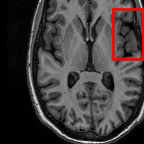}
\end{subfigure} &
\begin{subfigure}[t]{0.125\textwidth}
    \centering
    \includegraphics[width=\linewidth]{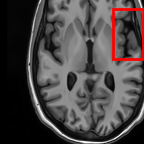}
\end{subfigure} & 
\begin{subfigure}[t]{0.125\textwidth}
    \centering
    \includegraphics[width=\linewidth]{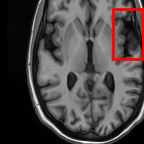}
\end{subfigure} & 
\begin{subfigure}[t]{0.125\textwidth}
    \centering
    \includegraphics[width=\linewidth]{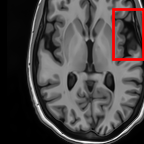}
\end{subfigure} & 
\begin{subfigure}[t]{0.125\textwidth}
    \centering
    \includegraphics[width=\linewidth]{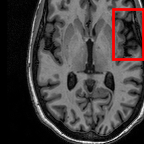}
\end{subfigure} & 
\begin{subfigure}[t]{0.125\textwidth}
    \centering
    \includegraphics[width=\linewidth]{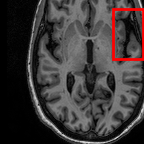}
    \vspace{-2em}
\end{subfigure}\\
&
\begin{subfigure}[t]{0.125\textwidth}
    \centering
    \includegraphics[width=\linewidth]{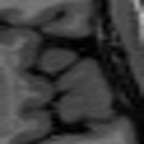}
    \vspace{-2em}
\end{subfigure} &
\begin{subfigure}[t]{0.125\textwidth}
    \centering
    \includegraphics[width=\linewidth]{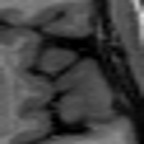}
    \vspace{-2em}
    \caption{\tiny{27.37/0.8465}}
\end{subfigure} &
\begin{subfigure}[t]{0.125\textwidth}
    \centering
    \includegraphics[width=\linewidth]{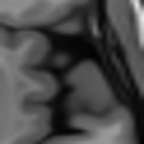}
    \vspace{-2em}
    \caption{\tiny{32.34/0.9441}}
\end{subfigure} &
\begin{subfigure}[t]{0.125\textwidth}
    \centering
    \includegraphics[width=\linewidth]{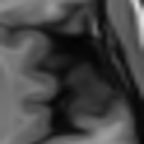}
    \vspace{-2em}
    \caption{\tiny{32.72/0.9436}}
\end{subfigure} &
\begin{subfigure}[t]{0.125\textwidth}
    \centering
    \includegraphics[width=\linewidth]{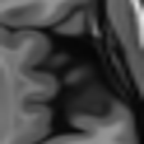}
    \vspace{-2em}
    \caption{\tiny{{\bf34.11}/{\bf0.9607}}}
\end{subfigure} &
\begin{subfigure}[t]{0.125\textwidth}
    \centering
    \includegraphics[width=\linewidth]{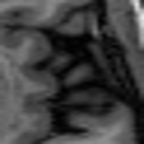}
    \vspace{-2em}
    \caption{\tiny{PSNR(dB)/SSIM}}
\end{subfigure} &
\begin{subfigure}[t]{0.125\textwidth}
    \centering
    \includegraphics[width=\linewidth]{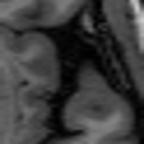}
\end{subfigure}\\

\hline
\multirow{4}{0.5em}{8} &

\begin{subfigure}[t]{0.125\textwidth}
    \centering
    \includegraphics[width=\linewidth]{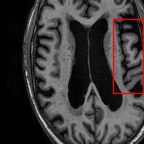}
\end{subfigure} &
\begin{subfigure}[t]{0.125\textwidth}
    \centering
    \includegraphics[width=\linewidth]{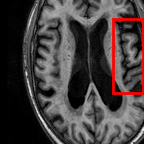}
\end{subfigure} &
\begin{subfigure}[t]{0.125\textwidth}
    \centering
    \includegraphics[width=\linewidth]{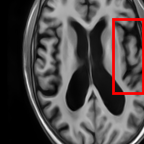}
\end{subfigure} & 
\begin{subfigure}[t]{0.125\textwidth}
    \centering
    \includegraphics[width=\linewidth]{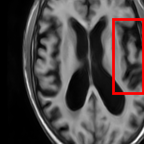}
\end{subfigure} & 
\begin{subfigure}[t]{0.125\textwidth}
    \centering
    \includegraphics[width=\linewidth]{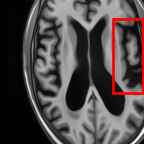}
\end{subfigure} & 
\begin{subfigure}[t]{0.125\textwidth}
    \centering
    \includegraphics[width=\linewidth]{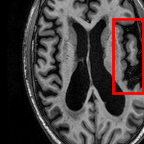}
\end{subfigure} & 
\begin{subfigure}[t]{0.125\textwidth}
    \centering
    \includegraphics[width=\linewidth]{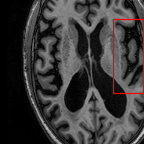}
\end{subfigure}\\
&
\begin{subfigure}[t]{0.125\textwidth}
    \centering
    \includegraphics[width=\linewidth]{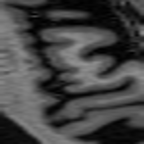}
\end{subfigure} &
\begin{subfigure}[t]{0.125\textwidth}
    \centering
    \includegraphics[width=\linewidth]{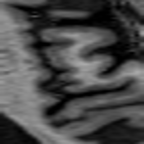}
    \vspace{-2em}
    \caption{\tiny{25.51/0.7681}}
\end{subfigure} &
\begin{subfigure}[t]{0.125\textwidth}
    \centering
    \includegraphics[width=\linewidth]{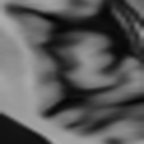}
    \vspace{-2em}
    \caption{\tiny{28.29/0.8205}}
\end{subfigure} &
\begin{subfigure}[t]{0.125\textwidth}
    \centering
    \includegraphics[width=\linewidth]{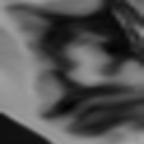}
    \vspace{-2em}
    \caption{\tiny{29.51/0.8824}}
\end{subfigure} &
\begin{subfigure}[t]{0.125\textwidth}
    \centering
    \includegraphics[width=\linewidth]{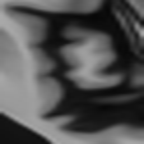}
    \vspace{-2em}
    \caption{\tiny{{\bf31.87}/{\bf0.9249}}}
\end{subfigure} &
\begin{subfigure}[t]{0.125\textwidth}
    \centering
    \includegraphics[width=\linewidth]{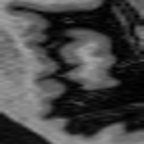}
    \vspace{-2em}
    \caption{\tiny{PSNR(dB)/SSIM}}
\end{subfigure} &
\begin{subfigure}[t]{0.125\textwidth}
    \centering
    \includegraphics[width=\linewidth]{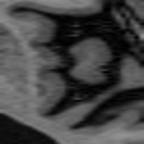}
\end{subfigure}\\

\hline

\end{tabularx}
\vspace{-1em}
\caption{Visual comparisons of slice interpolation approaches. For 4x sparsity, the second of three interpolated MR slices is presented. For 8x sparsity, the third of seven interpolated slices is presented.}\label{fig:Comparison}
\end{figure*}

\subsection{Visual Comparisons}
In Fig. \ref{fig:Comparison}, we present the observed slices $I_{\downarrow k}^z$ and $I_{\downarrow k}^{z+1}$ along with the interpolated slices produced by different methods. Specifically we demonstrate the second of three interpolated MR slices for 4x sparsity, and the third of seven interpolated slices for 8x sparsity. We highlight the region where the anatomical structures significantly change compared to the observed slices $I^{z}_{\downarrow k}$ and $I^{z+1}_{\downarrow k}$. We observe that although 2D CNN has comparable performance in terms of PSNR and SSIM, it tends to produce false anatomical structures in the zoomed regions. 3D CNN is able to resolve more accurate details. However, the improvement is quite limited, which we attribute to the fact that 3D CNN requires more training MR volumes in order to generalize and has smaller receptive field due to patch-based training. Our method benefits from the large receptive field of 2D CNN and two-view fusion, which not only produces sharper images, but also correctly estimates brain anatomy. The sharp and accurate estimation is crucial in clinical applications such as diagnosing Alzheimer's Disease by brain volume estimation.

\newcolumntype{Y}{>{\centering\arraybackslash}X}
\begin{figure*}[t!]
\captionsetup[subfigure]{labelformat=empty}
\begin{tabularx}{\textwidth}{ |c |*{5}{Y}| }
\hline

Sparsity& LI & 2D CNN & 3D CNN & Ours & GT \\
\hline
\multirow{2}{0.5em}{4} &
\begin{subfigure}[t]{0.125\textwidth}
    \centering
    \includegraphics[width=\linewidth]{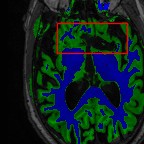}
    \vspace{-2em}
    \caption{\tiny{0.6787/0.7972}}
\end{subfigure} &
\begin{subfigure}[t]{0.125\textwidth}
    \centering
    \includegraphics[width=\linewidth]{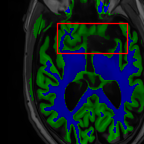}
    \vspace{-2em}
    \caption{\tiny{0.8143/0.8776}}
\end{subfigure} &
\begin{subfigure}[t]{0.125\textwidth}
    \centering
    \includegraphics[width=\linewidth]{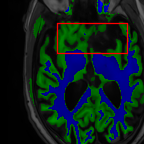}
    \vspace{-2em}
    \caption{\tiny{0.8190/0.8714}}
\end{subfigure} &
\begin{subfigure}[t]{0.125\textwidth}
    \centering
    \includegraphics[width=\linewidth]{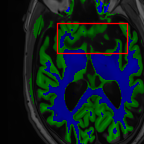}
    \vspace{-2em}
    \caption{\tiny{{\bf0.8664}/{\bf0.9085}}}
\end{subfigure} &
\begin{subfigure}[t]{0.125\textwidth}
    \centering
    \includegraphics[width=\linewidth]{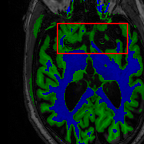}
    \vspace{-2em}
    \caption{\tiny{GM/WM}}
\end{subfigure}\\ 
&
\begin{subfigure}[t]{0.125\textwidth}
    \centering
    \includegraphics[width=\linewidth]{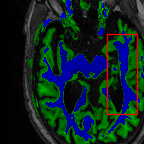}
    \vspace{-2em}
    \caption{\tiny{0.6808/0.7161}}
\end{subfigure}&
\begin{subfigure}[t]{0.125\textwidth}
    \centering
    \includegraphics[width=\linewidth]{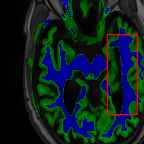}
    \vspace{-2em}
    \caption{\tiny{0.8103/0.8631}}
\end{subfigure}&
\begin{subfigure}[t]{0.125\textwidth}
    \centering
    \includegraphics[width=\linewidth]{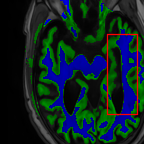}
    \vspace{-2em}
    \caption{\tiny{0.7950/0.8606}}
\end{subfigure}&
\begin{subfigure}[t]{0.125\textwidth}
    \centering
    \includegraphics[width=\linewidth]{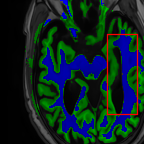}
    \vspace{-2em}
    \caption{\tiny{{\bf0.8598}/{\bf0.9115}}}
\end{subfigure}&
\begin{subfigure}[t]{0.125\textwidth}
    \centering
    \includegraphics[width=\linewidth]{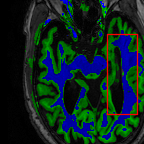}
    \vspace{-2em}
    \caption{\tiny{GM/WM}}
\end{subfigure}\\  

\hline
\multirow{2}{0.5em}{8} &
\begin{subfigure}[t]{0.125\textwidth}
    \centering
    \includegraphics[width=\linewidth]{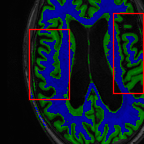}
    \vspace{-2em}
    \caption{\tiny{0.5139/0.7240}}
\end{subfigure} &
\begin{subfigure}[t]{0.125\textwidth}
    \centering
    \includegraphics[width=\linewidth]{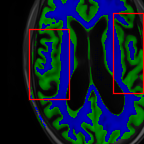}
    \vspace{-2em}
    \caption{\tiny{0.6619/0.8224}}
\end{subfigure} & 
\begin{subfigure}[t]{0.125\textwidth}
    \centering
    \includegraphics[width=\linewidth]{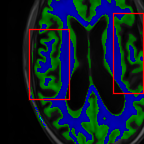}
    \vspace{-2em}
    \caption{\tiny{0.6878/0.8584}}
\end{subfigure} & 
\begin{subfigure}[t]{0.125\textwidth}
    \centering
    \includegraphics[width=\linewidth]{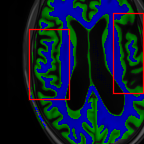}
    \vspace{-2em}
    \caption{\tiny{{\bf0.7798}/{\bf0.8853}}}
\end{subfigure} & 
\begin{subfigure}[t]{0.125\textwidth}
    \centering
    \includegraphics[width=\linewidth]{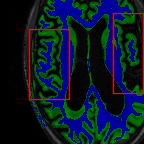}
    \vspace{-2em}
    \caption{\tiny{GM/WM}}
\end{subfigure}\\
&
\begin{subfigure}[t]{0.125\textwidth}
    \centering
    \includegraphics[width=\linewidth]{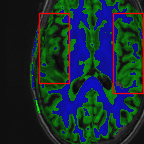}
    \vspace{-2em}
    \caption{\tiny{0.5910/0.6947}}
\end{subfigure} &
\begin{subfigure}[t]{0.125\textwidth}
    \centering
    \includegraphics[width=\linewidth]{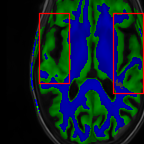}
    \vspace{-2em}
    \caption{\tiny{0.6516/0.8021}}
\end{subfigure} &
\begin{subfigure}[t]{0.125\textwidth}
    \centering
    \includegraphics[width=\linewidth]{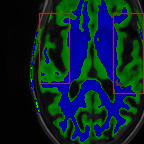}
    \vspace{-2em}
    \caption{\tiny{0.6507/0.8186}}
\end{subfigure} &
\begin{subfigure}[t]{0.125\textwidth}
    \centering
    \includegraphics[width=\linewidth]{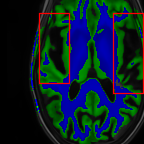}
    \vspace{-2em}
    \caption{\tiny{{\bf0.7471}/{\bf0.8540}}}
\end{subfigure} &
\begin{subfigure}[t]{0.125\textwidth}
    \centering
    \includegraphics[width=\linewidth]{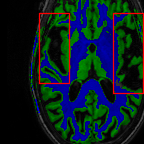}
    \vspace{-2em}
    \caption{\tiny{GM/WM}}
\end{subfigure}\\
\hline
\end{tabularx}
\vspace{-1em}
\caption{Visual comparison of gray matter (Green)/white matter (Blue) segmentation over different methods, with respective DICE scores listed under the images.}\label{fig:Seg}
\end{figure*}

In Fig. \ref{fig:Seg}, we demonstrate the advantage of the proposed method in brain matter segmentation. It is clear that although 2D and 3D CNN generates visually plausible interpolation as presented in Fig. \ref{fig:Comparison}, the brain matters are easily misclassified due to incorrect anatomical structures and blurred details.


\subsection{Ablation study} \label{sec:ablation_study}

In this section, based on 4x sparsity, we evaluate the effectiveness of each proposed components. The following settings are considered:
\begin{itemize}
    \item MSR$^{n}_{sag}$: Slice interpolation based on only sagittal view MSR. We consider number of input slices $n={1,3}$.
    \item MSR$^{n}_{cor}$: Slice interpolation based on only coronal view MSR. We consider number of input slices $n={1,3}$.
    \item Fused: Slice interpolation with fusion network. Inputs to the network are MSR$^{3}_{sag}$ and MSR$^{3}_{cor}$.
    \item Refined: The proposed full pipeline.
\end{itemize}

\newcolumntype{Y}{>{\centering\arraybackslash}X}
\begin{table*}[t!]
\begin{tabularx}{\textwidth}{ |c| *{2}{Y|} }
\hline

\hline
Stage & PSNR (dB) & SSIM  \\
\hline
baseline 2D CNN & 31.24&0.9313\\
baseline 3D CNN & 31.34&0.9292\\
\hline
MSR$^{1}_{sag}$ & 30.28 & 0.9129 \\
MSR$^{1}_{cor}$ & 30.56 & 0.9178  \\
\hline
MSR$^{3}_{sag}$ & 31.43 & 0.9314 \\
MSR$^{3}_{cor}$ & 31.61 & 0.9339  \\
Fused & \underline{32.02} & \underline{0.9413}\\
Refined & {\bf 32.22} & {\bf 0.9441} \\ 
\hline
\end{tabularx}
\vspace{2pt}
\caption{Quantitative ablation study. Baseline numbers are also included for comparison. The best results are in {\bf bold} and the second best \underline{underlined}.}\label{tab:ablation}
\vspace{-0.5cm}
\end{table*}

\begin{figure*}[t]
\captionsetup[subfigure]{labelformat=empty}
\def\arraystretch{0.1}
\begin{tabularx}{\textwidth}{ *{4}{Y} }
    \begin{subfigure}[t]{0.25\textwidth}
        \caption{GT}
        \centering
        \includegraphics[width=\linewidth,height=0.8\linewidth]{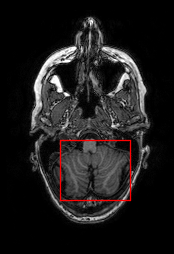}
    \end{subfigure} &
    \begin{subfigure}[t]{0.25\textwidth}
        \caption{MSR$^{1}_{sag}$}
        \centering
        \includegraphics[width=\linewidth,height=0.8\linewidth]{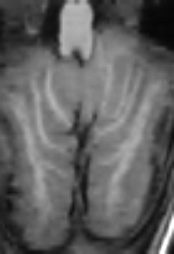}
    \end{subfigure} &
    \begin{subfigure}[t]{0.25\textwidth}
        \caption{MSR$^{3}_{sag}$}
        \centering
        \includegraphics[width=\linewidth,height=0.8\linewidth]{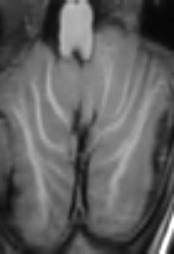}
    \end{subfigure} &
    \begin{subfigure}[t]{0.25\textwidth}
        \caption{Fused}
        \centering
        \includegraphics[width=\linewidth,height=0.8\linewidth]{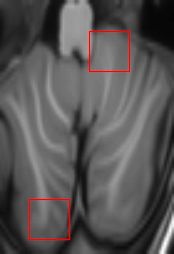}
    \end{subfigure} \\
    \begin{subfigure}[t]{0.25\textwidth}
        \caption{GT (zoomed)}
        \centering
        \includegraphics[width=\linewidth,height=0.8\linewidth]{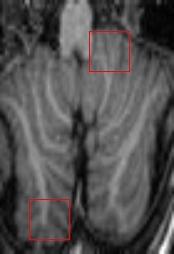}
    \end{subfigure} &
    \begin{subfigure}[t]{0.25\textwidth}
        \caption{MSR$^{1}_{cor}$}
        \centering
        \includegraphics[width=\linewidth,height=0.8\linewidth]{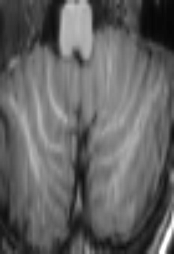}
    \end{subfigure} &
    \begin{subfigure}[t]{0.25\textwidth}
        \caption{MSR$^{3}_{cor}$}
        \centering
        \includegraphics[width=\linewidth,height=0.8\linewidth]{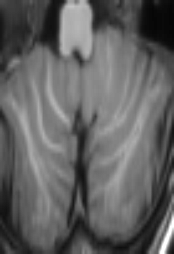}
    \end{subfigure} &
    \begin{subfigure}[t]{0.25\textwidth}
        \caption{Refined}
        \centering
        \includegraphics[width=\linewidth,height=0.8\linewidth]{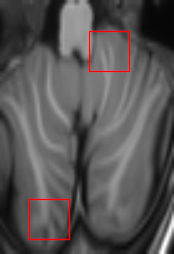}
    \end{subfigure}
\end{tabularx}
\vspace{-1em}
\caption{Visual comparison for the proposed components.
}\label{fig:ablation}
\end{figure*}
From Table \ref{tab:ablation}, it is clear that each proposed component improves the quality of slice interpolation. Notice that even without fusion and refinement, the axial slices interpolated by MSR$_{sag}^3$ and MSR$_{cor}^3$ are already better than the baseline 2D/3D CNNs.

Visual comparisons are shown in Fig. \ref{fig:ablation}, where we select a challenging slice with abundant anatomical details. From Fig. \ref{fig:ablation}, it is clear that marginally super-resolving axial slices from coronal and sagittal views leads to noticeable horizontal (MSR$^{n}_{sag}$) and vertical (MSR$^{n}_{cor}$) artifacts. Furthermore, some small details are better resolved by MSR$^{3}_{sag}$, while others are better resolved by MSR$^{3}_{cor}$. The fusion network combines the features from MSR$^{3}_{sag}$ and MSR$^{3}_{cor}$, which effectively reduces inconsistency. With the additional axial information, the fused slice is then further improved by the refinement network.

In addition to $L_1$ loss, we also experiment on GAN loss at refinement stage. However, we find that GAN tends to generate fake anatomical details, which is undesirable in medical applications.

\section{Conclusion}

In this work, we proposed a multi-stage 2D CNN framework called deep slice interpolation. This framework allows us to recover missing slices with high quality, even when the distance between observed slices are sparsely sampled. We evaluated our approach on a large ADNI dataset, demonstrating that our method outperforms possible 2D/3D CNN baselines both visually and quantitatively. Furthermore, we have illustrated that the MR slices estimated by the proposed method have superior segmentation accuracy. In the future, we plan to investigate the potential application of the proposed framework on real screening MRI which often have a very low slice density. 


\bibliographystyle{splncs}
\bibliography{main}
\end{document}